\documentclass[useAMS,usenatbib,usegraphicx]{mn2e}
\def\be{\begin{equation}}
\def\ee{\end{equation}}

\begin{document}

\title{Signatures of QCD Phase Transition in a Newborn Compact Star} 
\author[K.~W.~Wong and 
M.~-C.~Chu]{K.~-W.~Wong$^{1,2}$\thanks{E-mail:kwwong@virginia.edu;} and 
M.~-C.~Chu$^1$\thanks{E-mail:mcchu@phy.cuhk.edu.hk}\\
$^{1}$Department of Physics, The Chinese University of 
Hong Kong,  Shatin, New Territories, Hong Kong, China \\
$^{2}$Department of Astronomy, University of Virginia, Charlottesville, 
Virginia}
 
\date{Accepted 2004 March 17. Received 2004 March 2; in original form 2003 
November 25}

\pagerange{\pageref{firstpage}--\pageref{lastpage}} \pubyear{2004}

\maketitle

\label{firstpage}

\begin{abstract}
We study the scenario that a new born strange quark star cools to 
the Quantum ChromoDynamics (QCD) phase transition temperature and converts 
to a neutron star, and we calculate the evolution of temperature and 
luminosity of the compact star.  We argue that the conversion energy 
released can be of the order $10^{53}$ erg.  
We also propose that a second neutrino burst will be emitted at the completion
of this phase transition.   

\end{abstract} 

\begin{keywords}
stars: neutron - neutrino - dense matter.
\end{keywords}

\section{Introduction}
 
The Bodmer-Witten proposal that symmetric deconfined u,d,s-quark matter 
may be the absolute ground state of matter and forms the so-called strange
stars  \citep{Bodmer,Witten} has aroused much interest, and the properties  
of strange stars have been widely studied  
\citep{Alcock,Cheng98a,Farhi,Glendenning,Haensel}.  
An important question is whether the observed compact stars are neutron 
stars  or strange stars.  One possibility to distinguish the two is to study
their cooling curves, which are significantly different 
\citep{Weber,Ng,Usov1998,Usov2001,Usov2002}. 

In this paper, we study the effects of the Quantum ChromoDynamics (QCD) 
phase transition 
on the cooling of a compact star and possible signatures of the quark phase.
Regardless of the validity of the Bodmer-Witten proposal, 
the formation of quark-gluon plasma should be favoured in high 
temperature and density \citep{Shuryak}; 
we therefore suggest that a strange star may be formed just 
after a supernova explosion, in which both conditions are 
satisfied \citep{Benvenuto}. Because the initial temperature is so high 
\citep{Petschek} $T_i \sim40 {\rm MeV} $, the initial compact star is 
likely to be a bare strange star \citep{Usov2001}. 
When it cools down to the phase transition temperature $T_p$, 
the quark matter becomes energetically unstable compared to nuclear 
matter,  and the strange star will convert to a neutron star.  
The conversion energy
released during this QCD phase transition can be of the order $10^{53}$ erg. 
The temperature drops drastically at the completion of the phase 
transition, which is accompanied by 
the emission of a second neutrino burst due to the higher neutrino
emissivity of neutron matter. 

In Section 2, we will first present the equation of state (EOS) that we 
use in the calculation for the quark phase.    
This is then followed by a discussion of the
stability properties of strange stars in Section 3.  Section 4 is
devoted to the study of the phase transition
scenario, which is fixed once the EOS's are chosen.  We then discuss
the cooling processes of strange stars in Section 5, and the results
of our calculation are presented in Section 6.  We summarize in Section 7.

\section{COLD EQUATION OF STATE FROM PERTURBATIVE QCD}
Lattice QCD calculation of $T_p$ is highly uncertain at high chemical
potential, but the latest results \citep{allton} indicate, though with
relatively large uncertainties at high chemical potential, that 
$T_p $ drops from its zero density value of 140 MeV to about 
50 MeV at 1.5 times nuclear matter density 
$\rho_0=0.17 \textrm{ fm}^{-3}$ and down to a few MeV for
density a few times $\rho_0$.  Some previous proto-neutron star evolution
calculations indeed show that it is feasible to reach the phase transition
in supernovae \citep{Benvenuto}.  While there are still large uncertainties 
in both high density QCD and the proto-neutron star evolution, we believe
it is worthwhile studying the possible consequences of the QCD phase
transition in supernovae.
We assume a constant $T_p$ in the star and present results for 
$T_p = 1, \ 10$ MeV for comparison.  We adopt the simple picture that 
matter at temperature above (below) $T_p$ is in the quark (hadronic) 
phase.

To study the properties of quark matter, various EOS's have been used. 
The MIT Bag model is most widely used due to its simple analytic form 
\citep{Alcock}. Here we follow Fraga $et$ $al.~$ \citep{Fraga} and use the 
EOS derived from perturbative QCD for cold, dense quark matter up to 
second order in the strong coupling constant $ \alpha_s$. 
$\alpha_s$ becomes small in the high density limit, 
with a value of about 0.4 in the relevant density regime.
It turns out that this EOS is very similar to the MIT Bag model EOS,
with an effective Bag constant \citep{Fraga}, and we would have obtained 
basically the same results using the latter.  None of our results 
in the cooling calculation depends on the validity of perturbative QCD in 
compact star regime.

All thermodynamic properties can be obtained from the thermodynamic potential
$\Omega (\mu)$, where $\mu$ is the chemical potential.
At zero quark mass limit, the number densities of 
u, d, s quarks are the same, and hence charge 
neutrality is automatically satisfied, without any need of 
electrons.  The zero temperature perturbative QCD thermodynamic potential 
has been calculated up to order $\alpha_s^2$ \citep{Freedman,Baluni} 
in the modified  minimal subtraction scheme \citep{Baluni}:
\be
\Omega(\mu)=- \frac{N_f\mu^4}{4 \pi^2} \left\{1-2{\tilde \alpha _s} -
\left[G+N_f\ln {\tilde \alpha _s} + \beta_0\ln
{\tilde \Lambda}\right]{\tilde \alpha _s}^2\right\}, 
\ee
where $\beta_0=11-2N_f/3$, $N_f$ is the number of quark flavors, 
$G={G}_{0}+N_f(\ln N_f-0.536)$, ${G}_{0}=10.374\pm 0.13$, 
${\tilde \alpha _s} \equiv \alpha _s /\pi$, ${\tilde \Lambda} \equiv 
\Lambda/\mu$, $\Lambda$ being the renormalization subtraction point, and
\begin{equation}
{\tilde \alpha_s} (\Lambda)= \frac{4}{\beta_0 u}
\left\{ 1-{\tilde \beta}_1\ln u 
+ {\tilde \beta}_1^2 \left[ \left( \ln u - \frac{1}{2} \right)^2 +
{\tilde \beta_2}
 -\frac{5}{4} \right] \right\} \ ,
\end{equation}
with $u=\ln (\Lambda^2/\Lambda_{\overline{MS}}^2)$, 
${\Lambda}_{\overline{MS}}=365$ MeV,
${\beta}_{1}=51-19{N}_{f}/3$, 
${\beta}_{2}=2857-325{N}_{f}^2/27$, 
${\tilde \beta}_1 = 2\beta _1/\beta_0^2u$ and 
${\tilde \beta}_2 = \beta _2 \beta_0 / 8 \beta_1^2$.
It is believed that 
${\tilde \Lambda}$ lies in the range between 2 and 3 \citep{Fraga}.
Both the first and second order terms decrease the pressure of the 
strange quark matter relative to the ideal gas. 
The pressure depends weakly on the strange quark mass $m_s$, changing
only by less than 5\% for $m_s$ up to 150 MeV \citep{sor}.  We will therefore
use the massless EOS in the calculation.  

\section{STABILITY OF STRANGE QUARK MATTER}
The structure of a static, non-rotating and 
spherically symmetric strange star can be calculated by solving the 
Tolman-Oppenheimer-Volkov (TOV) equations together with the EOS 
\citep{Glendenning}. 
A strange star can be stable even at zero temperature if its binding energy
is larger than that of a neutron star with the same baryonic mass,
which is indeed the case for  ${\tilde \Lambda}$  around 2.7  (see Table I),
for several commonly used neutron star EOS's \citep{Weber}.  We are however 
interested
in the possibility that strange quark matter is only stable for $T>T_p$,
and so we choose a ${\tilde \Lambda} < 2.7$, so that when the hot strange star
cools to low temperature, it will convert to a neutron star.
For ${\tilde \Lambda}=2.473$, the maximum 
gravitational mass is $1.516 M_{\odot}$ with a baryonic mass of $1.60 
M_{\odot}$ and radius $8.54$ km. We will use this set of parameters in the
calculation of the cooling behaviour because the maximum mass is close to 
observational data of compact stars.  In fact, we have also used other
values of ${\tilde \Lambda}$, and the cooling behavior is qualitatively 
similar, as long as the star undergoes a phase transition.

\section{PHASE TRANSITION FROM STRANGE STARS TO NEUTRON STARS}
It has long been suggested that strange stars can be formed from a 
phase transition of neutron stars to strange stars due to an abrupt 
increase in density \citep{Cheng96,Cheng98b}. 
However, from the theoretical point of 
view, formation of quark-gluon plasma is favoured when both temperature 
and chemical potential are high enough \citep{Shuryak}. 
We propose that strange stars are formed in supernovae where 
both the temperature and density are high, with initial temperature 
\citep{Petschek} $T_i \sim40 {\rm MeV} > T_p$.  The star then cools to 
$T_p$ and hadronizes into a neutron star containing 
ordinary baryons.  This is just the same scenario believed to occur
in ultra-relativistic heavy-ion collisions \citep{Shuryak}. 
If the baryonic mass $M_{B}$ is conserved during the 
phase transition, the total conversion energy 
$E_{\textrm{\scriptsize{conv}}}$ released is: 
\be
E_{\textrm{\scriptsize{conv}}}=[M_{G}({\rm SS})-M_{G}({\rm NS})]c^2,
\ee
where $M_{G}({\rm SS})$ and $M_{G}({\rm NS})$ are the gravitational 
masses of the strange star and neutron star respectively \citep{Bombaci}. 
Whether a phase transition can occur and how much energy is released 
depend on both the EOS's of quark matter and nuclear matter. 
We choose several commonly used neutron star EOS's, 
$M_{G}=1.40M_{\odot}$ \citep{Weber}, and the conversion 
energy for different ${\tilde \Lambda}$ are summarized in 
Table I.  For ${\tilde \Lambda}=2.473$, typically $10^{53}$ erg is released 
during the conversion process, which depends
only weakly on the nuclear matter EOS. 

\section{COOLING PROPERTIES}

The surface of a newborn strange star is so hot that all the materials, 
other than quark matter, are evaporated leaving the strange star 
nearly bare without any crust 
\citep{Usov2001}. Since the thermal conductivity of strange matter 
is high and the density profile of the strange star is very flat, we 
take the uniform temperature and density approximation. 
The strange star cools according to: 
\be
C_q\frac{dT}{dt}=-L_q ,
\ee
where $C_q$ is the total heat capacity of all the species in quark 
matter, and $L_q$ is the total luminosity of the star.
When the temperature drops to $T_p$, the star undergoes a phase 
transition releasing a conversion energy $E_{\textrm{\scriptsize{conv}}}$.

During the 
phase transition, we assume that the quark and neutron matter are distributed 
uniformly and calculate the luminosity of the mixed phase by the 
weighted average of those of the quark matter and the neutron 
matter~\citep{directURCA,sor}. When the strange star has converted completely 
to a neutron star, it then follows the standard 
cooling of a neutron star with an initial temperature of $T_p$.

The detailed thermal evolution is governed by
several energy transport equations.
We adopt a simple model that a 
neutron star has a uniform temperature core with high conductivity and 
two layers of crust, the inner crust and the outer crust, which transport 
heat not as effectively as the core or quark matter. The typical thickness 
of the crust is $\sim 10\%$ of the radius, and we can use the parallel-plane 
approximation to describe the thermal evolution of the inner crust. 
The thermal history of the inner crust can be described by a heat conduction
equation:
\begin{equation}
c_{\textrm{\scriptsize{crust}}}\frac{\partial 
T}{\partial t} = \frac{\partial}{\partial r}\left(
K\frac{\partial T}{\partial r} \right) - \epsilon_{\nu},
\end{equation}
where $c_{\textrm{\scriptsize{crust}}}$ is the specific heat of the inner
crust, $K$ is the effective thermal conductivity, and $\epsilon_{\nu}$ is
the neutrino emissivity. As a rule of thumb, the effective surface
temperature $T_e$ and the temperature at the interface of inner and outer
crust, $T_b$, are related by \citep{Gudmundsson}:
\begin{equation}
T_{b8}=1.288(T_{s6}^4/g_{s14})^{0.455},
\end{equation}
where $g_{s14}$ is the surface gravity in the unit of $10^{14}$ cm
$\textrm{s}^{-2}$, $T_{b8}$ is the temperature between the inner and
outer crusts in the unit of $10^8$ K, and $T_{s6}$ is the effective 
surface
temperature in the unit of $10^6$ K. The luminosity at the stellar surface,
$L_{\textrm{\scriptsize{surface}}}$, is equal to the heat flux at the
interface of inner and outer crusts:
\begin{equation}
-K\frac{\partial T}{\partial
r}=L_{\textrm{\scriptsize{surface}}}/(4\pi R^2),
\end{equation}
where $R$ is the radius of the star. The boundary condition at the
interface of the core and inner crust is:
\begin{equation}
C_{\textrm{\scriptsize{core}}}\frac{\partial T}{\partial
t}=-K\frac{\partial T}{\partial
r}A_{\textrm{\scriptsize{core}}}-L_{\nu}^{\textrm{\scriptsize{core}}},
\end{equation}
where $C_{\textrm{\scriptsize{core}}}$ is the total heat capacity of the
core,
$A_{\textrm{\scriptsize{core}}}$ is
the surface area of the core, and $L_{\nu}^{\textrm{\scriptsize{core}}}$ is
the total neutrino luminosity of the core.

\subsection{Heat capacity of quark stars}
The total heat capacity is the sum of the heat capacities of all 
species in the star. Without the effect of superfluidity, the quark 
matter can be considered as a free Fermi gas, with a specific heat 
\citep{Iwamoto} ${c}_{q}=2.5\times 10^{20}{\tilde \rho}^{2/3}{T}_{9}
\textrm{ erg cm}^{-3}\textrm{ K}^{-1}$,
where ${\tilde \rho}$ is the baryon density in units of $\rho _0$ and $T_9 
\equiv T/10^9$ K. In the superfluid state, the specific heat is modified as 
\citep{Horvath,Maxwell}: 
\be
c_{q}^{\textrm{\scriptsize{sf}}}=
3.15c_{q}  e^{-{1.76\over{\tilde T}}} \left[{2.5\over {\tilde T}}
-1.66  +3.64{\tilde T} \right], \    
        \textrm{for ${\tilde T}\leq 1$},
\ee
where ${\tilde T} = T/T_c$, $k_BT_c=\Delta /1.76$, 
and $\Delta$ is the energy gap in MeV.
It has been argued that for quark matter, even with unequal quark 
masses, in the Color-Flavor Locked (CFL) phase in which all the three 
flavors and colors are paired, quark matter is automatically charge neutral 
and no electrons are required \citep{Rajagopal}. However, for sufficiently 
large strange quark mass and the 
relatively low density regime near the star surface, the 2 color-flavor 
SuperConductor (2SC) phase is preferred. Therefore in a real 
strange star, electrons should be present. The contribution of 
electrons can be parametrized by the electron fraction $Y_e$ which 
depends on the model of strange stars. We choose $Y_e=0.001$ as a typical 
value. The specific heat capacity of electrons in the strange star  
is given by \citep{Ng} 
$c_{e}=1.7\times 10^{20} \left( Y_{e}{\tilde \rho} \right)^{2/3}T_{9}
\textrm{ erg cm}^{-3}\textrm{ K}^{-1} \ $, which
is unaffected by the superfluidity of quark 
matter. Hence it dominates the total heat capacity of the strange star 
when the temperature drops below $\sim T_c$.

\subsection{Luminosity of quark stars}
The total luminosity is the sum of all the energy emission mechanisms, 
including photon and neutrino emission. 
The dominating neutrino emission mechanism is the quark URCA process, 
with emissivity \citep{Iwamoto}:
\be
\epsilon_{d}\simeq 2.2\times 
10^{26}\alpha_{s} \left( {\tilde \rho} \right)Y_{e}^{1/3}T_{9}^{6}
\textrm{ erg cm}^{-3}\textrm{s}^{-1},
\ee
and we have chosen 
$\alpha_s=0.4$ as a constant value throughout the quark star. In 
the superfluid state, the neutrino emissivity is suppressed by a factor of 
$\exp(-\Delta /T)$.

It has been pointed out that the bare surface of a strange star is a 
powerful source of $e^+e^-$ pairs due to the strong surface electric field 
\citep{Usov1998}.  We adopt the $e^+e^-$ pair luminosity given in  
\citep{Usov2001} for our calculation.
We also include the thermal equilibrium and non-equilibrium blackbody
radiation in our calculation using standard treatment 
\citep{Alcock,Usov2001}; the contribution of the latter is small compared 
to other mechanisms at high temperature.
Once the temperature drops, the cooling is
dominated by the relatively low power non-equilibrium blackbody radiation,
as long as the star is still in the quark phase.

\subsection{Microphysics of the neutron star}
There are many different models of neutron star cooling. 
We adopt the one described by Ng \citep{Ng,Maxwell}. 
The neutrino emission mechanisms are the direct URCA processes 
\citep{directURCA}, the electron-proton Coulomb scattering in the crust 
\citep{Festa}, and the neutrino bremsstrahlung \citep{Ng}.
The surface luminosity will be of the blackbody radiation 
$L_{bb}=4\pi R^{2}\sigma T_s^{4}$,
with the 
effective surface temperature $T_s$.
The blackbody radiation will be the dominating cooling mechanism after 
neutrino emissions are switched off.

For the thermal conductivity of the inner crust $K$, we use a 
temperature dependent model \citep{Lindblom},
$K=2.8 \times 10^{21}/T_9 \textrm{ erg cm}^{-1}\textrm{ s}^{-1}\textrm{ 
K}^{-1}$.
The choice of $K$ will not be important after the epoch 
of thermal relaxation, which is of the order $10-100$ years. The 
temperature of the inner crust and the core will be uniform after that.
\\
When a strange star is born just after the stellar collapse, its 
temperature is very high, of the order $10^{11}$ K~\citep{Petschek}.  
${\tilde \Lambda}$ mainly affects the conversion energy, which affects 
the duration of the phase transition. The cooling mechanisms depend only 
weakly on ${\tilde \Lambda}$ while the cooling curves depend weakly 
on $T_i$.  We assume $T_i = 40$ MeV for our calculations. We choose a gap 
value $\Delta=100$ MeV \citep{Ng} to describe the superfluidity 
phase of quark 
matter, and we have checked that using $\Delta = 1$ MeV gives qualitatively
similar results \citep{sor}.

\section{Results}
The observables are the 
luminosity and the surface temperature at infinity, $L^{\infty}$ and 
$T_s^{\infty}$, which are related to the stellar surface values, $L$ and 
$T_s$ \citep{Tsuruta}: $T_s^{\infty}=e^{\phi_s}T_s$,
$L^{\infty}=e^{2\phi_s}L$,
where $e^{\phi_s}=\sqrt{1-2M_G/R}$ is the gravitational redshift at the 
stellar surface. The various cooling curves (solid lines) for $T_p = 1 \ (10)$
MeV are shown in left (right) panels of Fig.~1. 

For a large range of parameter values and nuclear matter EOS's, we obtain
a large energy released, of order $10^{53}$ erg.  The duration of this 
energy release depends
sensitively on $T_p$; a higher $T_p$ results in
a higher luminosity and shorter phase transition duration.
It can be as short as seconds for $T_p = 10$ MeV,
or as long as hundred thousands of seconds for $T_p = 1$ MeV.   
The surface temperature of the star drops rapidly (top panels of Fig.~1), 
reaching $10^7$~K already within the first second ($T_p = 10$ MeV) to first 
hundred thousand seconds ($T_p = 1$ MeV). 
The decrease in surface temperature is particularly drastic at the 
completion of the 
conversion, which is a unique feature of the phase transition not seen in
either a pure strange star (dotted curves) or a pure neutron star 
(dot-dashed curves).

The photon luminosity (middle panels in Fig.~1)
is initially dominated by the $e^+e^-$ pair emission
mechanism and is large in the quark phase due to the strong surface
electrostatic field.  Since the $e^+e^-$ pairs are not affected by
the superfluidity gap, the photon luminosity is hardly affected by
the gap values. The total energy radiated by  $e^+e^-$ pair emission is 
$4.24\times10^{50}(1.47\times10^{53})$~erg for $T_p=10(1)$~MeV.
Once the phase transition is completed, the surface
field of a neutron star is much weaker and this mechanism is turned off.
The photon luminosity therefore drops by over ten orders of magnitudes,
and this is more drastic for larger $T_p$, because the phase transition
occurs earlier and with shorter duration.  

A distinct second burst of neutrinos shows up in the neutrino luminosity
(bottom panels of Fig.~1), which accompanies the completion of the phase 
transition and arises because of the higher neutrino emissivity in neutron 
matter.  For $T_p = 10$ MeV, the neutrino flux rises 
by over ten orders of magnitudes within a small fraction of a second.
Note that we have not incorporated
a detailed transport calculation for the neutrinos, which results in a
broadening of the neutrino bursts we present here.  Indeed, for $T_p = 10$ MeV,
the peak neutrino intensity of about $10^{56}$ ergs$^{-1}$ lasts only for
$10^{-3}$ s, and these neutrinos will be spread out over a diffusion
time scale of ($\sim 1-10$ s) \citep{Petschek}, reducing the peak intensity 
by a factor of $10^3$.  The first and second neutrino peaks are likely rendered
indistinguishable by the relatively slow neutrino diffusion out of the dense
medium.  A much more careful treatment of the neutrino transport is clearly
needed here \citep{Liebendorfer}. 
However, if $T_p$ is as low as 1 MeV, the 
two bursts of similar flux can be separated by as long as $10^5$~s, which 
should be observable by modern neutrino observatories.  
The two neutrino bursts can in principle be distinquished also by
their energy spectra.  The first burst is emitted near the initial high
temperature of the newborn strange star, while the second burst is associated
with the phase transition temperature $T_p$, and therefore the second neutrino
burst has a softer energy spectrum.  The total energy 
radiated in neutrino is $3.15(1.27)\times10^{53}$~erg for $T_p=10(1)$~MeV.

This scenario of a second burst of neutrinos can be compared to two 
previous similar proposals \citep{Benvenuto,Aguilera}.
In our model, 
the burst is due to the phase transition from a quark star to a neutron star, 
which has a higher neutrino emissivity, whereas in previous proposals, the
second burst accompanies the phase transition from a neutron star to a quark
star.  In Benvenuto's theory, the phase transition is delayed by a few seconds
after the core bounce due to the presence of the neutrinos \citep{Benvenuto}.
In Aguilera $et$ $al.$'s theory \citep{Aguilera}, 
the burst is due to the initial trapping of neutrinos when the 
temperature is high and their sudden release when the quark star cools. 
If quark matter is not as stable as nuclear matter at low temperature, then
there should be yet another phase transition back to nuclear matter, which is
what we focus on, and the ``second'' neutrino burst we proposed is then the
``third'' neutrino burst.

If multiple neutrino bursts are observed, as may indeed be the case for 
the Kamiokande data for SN1987A \citep{Hirata}, whether the compact star 
changes from the quark phase to neutron phase (our model) 
or the other way around 
can be distinguished observationally in at least two ways. 
First, our model predicts that the cooling is much faster {\it before} 
the phase transition, but it will become slower {\it after} it.  
Second, the size of the post-phase-transition 
compact star, being a normal neutron star, would be larger in our model.

\section{Discussion and conclusion}
Based on the lattice QCD phase diagram, we propose that the 
new born compact star in a supernova is a strange star and it 
transforms to a neutron star when it cools down to a critical 
temperature $T_p$.  The conversion energy 
can be of the order $10^{53}$ erg and is adequate to supply energy for
GRB's.  The strange star cools rapidly due to neutrino and $e^+e^-$
emission, and its surface temperature drops drastically at the completion of
the conversion, when a second neutrino burst emerges due to the 
higher neutrino emissivity of neutron matter.
In our models, the phase transition is treated in a simplified manner.
Hydrodynamic calculation is needed for a detailed description of 
the process starting from a supernova explosion. 
Here we discuss semi-quantitatively the
signatures left by the quark to hadron phase transition,
if it occurs in supernovae.

This work is partially supported by 
a Hong Kong RGC Earmarked Grant CUHK4189/97P and a Chinese University
Direct Grant 2060105.  We thank Prof.~K.~S.~Cheng for useful discussion.


\newpage

\begin{table}
\caption{Total conversion energy $E_{\textrm{\scriptsize{conv}}}$
for various ${\tilde \Lambda}$. A neutron star 
gravitational mass $M_G({\rm NS})=1.4M_{\odot}$ is assumed, and 
the baryonic masses of strange stars are chosen to equal those of the 
neutron stars. The many-body approximation for HV, HFV, 
$\Lambda^{\textrm{\scriptsize{RBHF}}}_{\textrm{\scriptsize{BroB}}}+{\textrm{HFV}}$ and      
$\textrm{G}^{\textrm{\scriptsize{K240}}}_{\textrm{\scriptsize{M78}}}$ 
EOS's are relativistic Hartree, relativistic Hartree-Fock, relativistic 
Brueckner-Hartree-Fock + relativistic Hartree-Fock and relativistic 
Hartree respectively (Weber 1999).}
\label{Econv2}

\begin{tabular}{|c|c|c|c|c|}
\hline
EOS(NS) & $M_B/M_{\odot}$ & ${\tilde \Lambda}$ & 
$M_G({\rm SS})/M_{\odot}$ & $E_{\textrm{\scriptsize{conv}}}/10^{53}$ erg
\\\hline
    HV & 1.51 & 2.473  & 1.44 & $+0.72$
\\     &      & 2.600  & 1.41 & $+0.18$
\\     &      & 2.880  & 1.33 & $-1.25$
\\     &      & 3.000  & 1.29 & $-1.97$
\\\hline

   HFV   & 1.60 & 2.473  & 1.516 & $+2.08$
\\      &      & 2.600  & 1.478 & $+1.40$         
\\      &      & 2.880  & 1.400 & 0
\\      &      & 3.000  & 1.363 & $-0.66$
\\\hline
$\Lambda^{\textrm{\scriptsize{RBHF}}}_{\textrm{\scriptsize{BroB}}}+$HFV$$     
    & 1.62  & 2.473  &   /  &   /     
\\  &       & 2.600  & 1.50 & $+1.79$ 
\\  &       & 2.880  & 1.41 & $+0.18$
\\  &       & 3.000  & 1.38 & $-0.36$
\\\hline
$\textrm{G}^{\textrm{\scriptsize{K240}}}_{\textrm{\scriptsize{M78}}}$     
& 1.56 & 2.473  & 1.48 & $+1.43$ 
\\  &     & 2.600  & 1.45 & $+0.90$
\\  &     & 2.880  & 1.37 & $-0.54$ 
\\  &     & 3.000  & 1.33 & $-1.25$
\\\hline
\end{tabular}
\end{table}

\begin{figure}
\includegraphics[angle=0,width=8cm,angle=0]{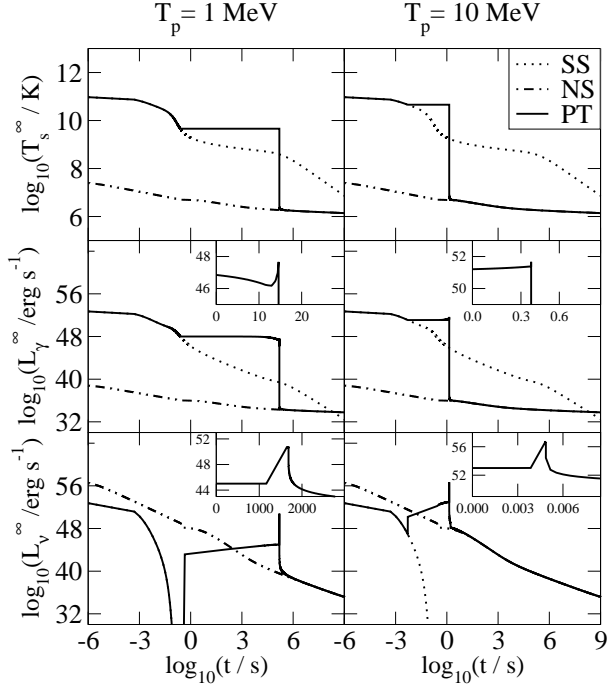}
\caption{Cooling curves corresponding to two different models. Left 
(right) panels are for models with small (large) phase transition 
temperature $T_p$ = 1 (10) MeV. The dotted (dashed) lines are the cooling 
curves of a pure strange (neutron) star without phase transition.  
The solid line represents the scenario with phase transition. 
$T_s^\infty, \ L_\gamma^\infty$ and $L_\nu^\infty$ 
denote surface temperature, photon luminosity and neutrino luminosity 
respectively.  $T_i = 40$ MeV and $\Delta$ = 
100 MeV are assumed. The small insets indicate the peak luminosities for 
models with phase transition (PT). The time axes of the small insets of 
the upper 
left, lower left, upper right and lower right start at $t$=155670, 
155400, 1 and 1.369 s respectively.} 
\end{figure}

\bsp

\label{lastpage}

\end{document}